# Avalanche effects in solid-phase epitaxial crystallization induced by light-ion irradiation

**N.E.B. Cowern[1*], B.J. Pawlak[2], and R. Gwilliam[3]**

[1]School of Electrical and Electronic Engineering, University of Newcastle upon Tyne, Newcastle upon Tyne, NE1 7RU, UK
[2]Globalfoundries, Kapeldreef 75, B-3001 Leuven, Belgium
[3]Surrey Ion Beam Centre, Nodus Laboratory, University of Surrey, Guildford, Surrey, GU2 7XH, UK.

Abstract

Solid-phase epitaxial crystallization of amorphous Si layers on a crystalline Si substrate during B-ion irradiation is investigated over the temperature range 293 – 573 K. Regrowth occurs at all measured temperatures, with activation energy ~ 0.07 eV and prefactor 2.36 nm / $10^{14}$ B cm$^{-2}$. The low activation energy suggests that free interstitial point defects are involved and the unusually large prefactor indicates that each interstitial crystallizes ~ $10^2$ host Si atoms. We propose that interstitial annihilation at the a/c interface sets off a shock-like process that drives c-Si island nucleation and growth until terminated by misfit strain.

Solid-phase epitaxial regrowth (SPER) of ion beam amorphized layers is an important technology for creating single-crystal layers with non-equilibrium properties such as enhanced doping or high strain levels. Currently for example it is a key contender for creation of highly strained layers for mobility enhancement in sub-14 nm CMOS technology. Remarkably however, the microscopic processes involved in SPER remain obscure. The effect of strain on the SPER velocity suggests intriguingly that the process may not be continuous but mediated on two timescales by nucleation and growth of crystalline islands at the amorphous/crystalline (a/c) interface [1]. However, current understanding of the mechanisms involved is limited, making it difficult to predict and control interface properties and defect formation at the nanometer scale.

Under thermodynamic equilibrium conditions in silicon, SPER proceeds at an average (macroscopic) rate with a characteristic Arrhenius dependence on temperature activated with an energy of about 2.7 eV. Under ion irradiation, growth rates at lower temperatures can be dramatically accelerated with activation energies reported in the 0.2 – 0.4 eV range. This suggests either direct ion-beam effects on the amorphous/crystalline (a/c) interface (Ref [2] and references therein) or the involvement of crystal point defects in rate-limiting steps of irradiation-induced SPER [2-4].

High energy ion implantation experiments show significant beam-induced regrowth during the implantation process. This has been found true for a wide range of ion masses, from B to Au, at energies which leave the implanted species deep within the crystalline material beyond the a/c interface, and thus unavailable to interact chemically with the interface. Reported activation energies range from ~ 0.2 eV for light ions (B, C, O) implanted at low temperature to ~ 0.4 eV for heavy ions (Ge, Au) [3,4]. The transition to the higher of these activation energies has been attributed to dissociation of di-vacancies formed in the ion collision cascade [3], although more probably di-vacancy dissociation has an energy barrier ~ 1.7 eV and di-vacancy diffusion has an energy barrier ~1.2 eV [Ref. 5 and references therein]. Values in the range 0.2 – 0.4 eV in p-type c-Si are consistent with vacancy migration [6,7], but at 0.2 eV activation energy, jump rates based on reasonable migration entropy values are high enough to transport the relevant point defect over large distances during implantation. In such cases the activation energy for point-defect induced regrowth may be better interpreted as a difference between the energy cost of acting on the a/c interface causing regrowth, and the energy cost of reaching alternative sinks within the silicon crystal.

In this work we explore the regrowth of a-Si layers on c-Si substrates under conditions where the dominant point defects in the underlying c-Si are interstitial. A 95 nm a-Si layer is initially formed by 75 keV Ge implantation into a c-Si substrate to a dose of $1 \times 10^{15}/cm^2$. After amorphization, B is implanted at 20 keV to a dose of $3 \times 10^{15}/cm^2$. The chosen implantation energy places the deeper portion of the B implant in the c-Si region, creating an excess of interstitial point defects over vacancies in the c-Si region, while the corresponding 'vacancy-rich' region is embedded in the upper portion of the a-Si layer. B implants are done at temperatures ranging from 77 K to 573 K. As reported in recent work [8], during implantation the B redistributes within the c-Si region as a result of interstitial mediated diffusion, while at the same time the a/c interface moves significantly as a result of beam-induced regrowth. A pile-up peak of B occurs at the a/c interface position, creating an accurate quantitative marker of the interface movement caused by regrowth. Here we focus in detail on the beam-induced regrowth process.

Fig. 1 shows B profiles measured by SIMS after implantation at a range of temperatures into the a-Si/c-Si bilayer structures described above [8]. Surprisingly, substantial beam-induced regrowth occurs at quite low temperatures – for example, the difference in a/c interface position for B implants at 293 K and 423 K (room temperature and 150°C) is about 8 nm. In contrast, based on earlier ion-beam studies at high implant energies, almost no movement would have been expected at these temperatures [4].

The position of the a/c interface as determined from the pile-up positions in Fig. 1 is plotted as a function of $1/kT$ in Fig. 2. The depth values for temperatures from 293 K upwards (on the left side of the figure) are determined with a relative accuracy of ± 1 nm. The depth uncertainty at 77 K is estimated at ± 2 nm, as the dip and peak present at the other temperatures is replaced by a discontinuity in the slope of the curve (visible by comparison with the 293 K curve in Fig. 1), and the B depth profile is also shifted due to the presence of a thin ice film on the sample forming on cooling prior to implantation, and had to be shifted back during the analysis in order to overlap the other SIMS curves. The solid curve plots the equation $x(T) = x_0 - r\phi \exp(-E_a/kT)$, where $x(T)$ is the a/c interface depth measured after B implantation at temperature $T$, $x_0$ is the initial a/c interface depth, $r$ is the prefactor for the regrowth distance per implanted dose, and $E_a$ is the activation energy for the beam-induced regrowth in our study. The parameters $x_0$, $r$ and $E_a$ are fitted to the data points at room temperature and above, for which the a/c interface position is sharply defined in the SIMS profiles. Although the 77 K data point is not included in this analysis it turns out to be right on the fitted curve, marking the a/c interface depth generated by the initial Ge implant.

Fig. 3 shows an Arrhenius plot of the shift in interface position, $\Delta x = x_0 - x(T)$. As expected from the above analysis the experimental data (symbols) fall on a straight line consistent with a single activation energy in this temperature range under the conditions of our experiment. The plotted straight line represents the fit described in the previous paragraph, now expressed as $\Delta x = r\phi \exp(-E_a/kT)$. This result differs from previously reported SPER data obtained after high energy light ion implantation (B, C, O), which showed the presence of multiple activation energies. Our extracted activation energy is 0.07 eV, nearly three times lower than the lowest previously reported activation energy for ion beam-induced SPER [3].

In our study the dominant point defects generated in the c-Si region are of interstitial, not vacancy, type. Obvious candidates are the free self-interstitial, *I*, whose diffusion in p-type Si under irradiation conditions is thought to be barrier free at low temperature [7,9], and the free boron-interstitial complex, *BI*, which is known to migrate at RT during ion beam irradiation [10,11]. We know from Fig. 1 and previously published results [8] that *BI* migrates to the a/c interface and accumulates there at high concentration during implantation at room temperature and above, saturating the interface with ~ $10^{-1}$ of a monolayer of B. Conceivably these atoms might catalyse re-bonding in their locality, enabling Si atoms in the amorphous phase close to the interface to incorporate into the Si crystal. However, it is implausible that a purely catalytic effect could reduce the barrier at the rate-limiting step from its thermal equilibrium value of about 2.7 eV down to as small a value as 0.07 eV.

The efficiency with which B implantation induces SPER is remarkable. Our measured dose-related regrowth prefactor is 2.36 nm / $10^{14}$ cm$^{-2}$, implying in the high-temperature limit that 118 Si atoms are crystallized from the amorphous phase for every implanted B ion. Since only about 15% of the implant stops in the c-Si, this implies the crystallization of ~ 800 atoms per B ion stopped in the c-Si (which we denote '/B'). This greatly exceeds the number of point defects generated in this region by the implant: SRIM simulations suggest about 300 atomic displacements/B, however most of the generated $I$ and $V$ migrate and recombine, either via the reaction $I + V \to 0$ or the two-step 'dopant-assisted recombination' reaction

$$I + B \to BI \quad (1)$$
$$BI + V \to B, \quad (2)$$

where $B$ represents substitutional B. Reaction (1) is very fast owing to the high B concentration present during most of the implantation period. As there is no significant barrier to reaction (1) [12], the interstitial travels a distance $(4\pi a C_B)^{-1/2} \approx 1$ nm before reacting with $B$ and the second reaction returns $BI$ to $B$ within a small distance, as $V$ is a much faster diffuser than $BI$. Thus, as $BI$ diffuses it rapidly recombines the excess $I$ and $V$ in the bulk c-Si region, taking with it a single interstitial from this region. As a result, the only interstitial point defects able to reach and potentially interact with the a/c interface are self-interstitials generated within the first 1-2 nm from the a/c interface (10 - 20 /B based on SRIM calculations), and BI from larger distances ($\leq$ 1 /B). For example, if 15 $I$/B and ~1 $BI$/B reach the a/c interface, then to account for our large growth prefactor, on average each interstitial defect must crystallize $\approx$50 Si atoms.

It is of great interest to discover what process induces SPER with such a small energy barrier and high crystallization yield. It seems natural that interstitial atoms would promote beam-induced SPER as their capture into substitutional c-Si sites at the a/c interface compensates tensile strain that develops as the c-Si phase encroaches into the less dense a-Si phase. Moreover, both $I$ and $BI$ are mobile at room temperature in p-type c-Si [7,9]. Since the volume mismatch $\eta$ between crystalline and amorphous Si is $\approx$2%, full volume compensation could be achieved if the capture of one interstitial initiates the recrystallization of $N = 1/\eta \approx 50$ Si atoms. This figure matches our estimated value based on the experimentally measured growth prefactor, suggesting that the volume compensation argument has validity. However, it remains a considerable challenge to find a mechanism by which one interstitial can initiate the crystallization of so many a-Si atoms.

We first consider two key consequences of the annihilation of an interstitial at the a/c interface. First, the process is exothermic, releasing an amount of energy $E_{ex} = E_{fI} + (E_{\alpha 1} - E_{\alpha 2})$, where $E_{fI} \sim 4$ eV is the interstitial formation energy in c-Si and $E_{\alpha 1}$, $E_{\alpha 2}$ are the total energies of the a/c interface before and after capture of the interstitial, respectively. $E_{\alpha 2}$ may exceed $E_{\alpha 1}$ by a few tenths of an eV owing to the addition of compressive strain energy as the interstitial is incorporated, as discussed below, but this still leaves $E_{ex} \approx E_{fI}$ to a good approximation. Thus $E_{ex}$ exceeds the experimentally reported recrystallization barrier $E_b \sim 2.7$ eV (actually somewhat lower in the p-type Si produced by B implantation). This enables the interstitial to incorporate as a substitutional atom at the interface, releasing at least 1 eV of vibrational energy. Our experimentally determined 0.07 eV activation energy can now be interpreted as the forward reaction barrier for interstitial incorporation. Second, when the interstitial atom moves to its new substitutional site, its volume increases from a fraction of one atomic volume to a full atomic volume. Both the energy release and the volume change take place in a fraction of a picosecond – an adiabatic event.

The exact way in which this leads to crystallization of a large number of atoms is open to discussion. One possibility is that the > 1 eV energy release and associated adiabatic expansion launches a compression shock, a portion of which is travels as a surface acoustic wave (SAW) along the a/c interface as a result of the impedance mismatch between c-Si and a-Si [13-15]. This concept is similar to that of nonlinear SAWs generated by sub-nanometer laser pulses [16], but here the driver is a more intense and localized vibrational energy release. According to our proposal, the shock drives explosive crystallization (ER), enabling a c-Si step to propagate along with the compression part of the wave, feeding more vibrational energy into the wave as a-Si converts to c-Si with the release of 0.1 eV/atom [17].

The self-sustaining nature of such a process has features in common with thermally driven ER during laser [18] or plasma jet [19] annealing, but the mechanism is fundamentally different. Thermal ER is believed to be initiated by contact between liquid Si and a crystalline substrate or microcrystalline seed which is also at high temperature. It is sustained by the latent heat of melt crystallization, and propagates sustainably with velocity in the region of ~ m/s directly into the a-Si volume until the material is consumed or the crystallization front enters a region where the substrate temperature is too low [19]. Such a mechanism could not apply in our case as we have an initial point-like source of energy which would dissipate ineffectively into three dimensions if only random vibrational energy were involved. In our proposed model, the interface structure dynamically couples vibrational energy release with lateral crystallization, driving the growth of a c-Si island.

In this picture, each avalanche is self-limiting because, as the island grows, the volume mismatch between a-Si and c-Si ensures that initial compressive strain energy on capture of the interstitial reverses to tensile, located primarily at step edges. The island stops growing when the volume contributed by the initially captured atom has become fully strain-compensated. The next time an interstitial is captured by the interface a new island nucleates and the process repeats itself. This model elegantly makes a connection between the number of atoms crystallized and the volume mismatch between a-Si and c-Si: one interstitial annihilation event crystallizes an average of $N = 1/\eta \approx 50$ atoms. Thus we have a self-limiting process that can be repeated without limit, with a rate determined by the arrival flux of interstitials during irradiation, leading to an average regrowth velocity prefactor that is in accord with experiment.

The high crystallization rate in our experiment may be significantly assisted by the presence of 0.1 monolayer of B at the interface [6], which could reduce the energy barrier to crystallization thus facilitating the process we have described. Further low-energy implantation studies using a range of ion species may help elucidate this potential contribution. The problem may also be accessible to molecular dynamics simulations.

In conclusion, we have conducted experiments using B ion implantation through an a-Si layer into a c-Si substrate, at temperatures ranging from 77 K to 573 K. The implantation energy is chosen so that the interstitial-rich implant tail lies in the c-Si region. This treatment causes substantial SPER during implantation, even at room temperature. The activation energy for the process, 0.07 eV, is the lowest yet measured for ion-beam induced SPER, and is consistent with an interstitial point-defect mechanism. The yield in terms of interface motion per implanted ion suggests that, on average, each interstitial initiates the incorporation of a large number of 'amorphous' Si atoms into the crystalline region. Strain and energy arguments linked to island nucleation and growth suggest the number should be inversely related to the volume mismatch between the amorphous and crystalline phases, i.e. ~ 50 in the case of silicon. This number is consistent with our regrowth data. Our analysis points to complex structural dynamics in the annealing of nanoscale amorphous/crystalline structures, which may lead to enhanced predictive capability for low-temperature SPER processing of three-dimensional nanoelectronic devices.

* Contact author e-mail: nick.cowern@ncl.ac.uk, tel.: +44.191.208.5636.

**References:**

1. N.G. Rudawski and K.S. Jones, Scripta Materialia **61**, 327 (2009).
2. B. Sklenard (PhD thesis, University of Grenoble, May 2014).
3. A. Kinomura, A. Chayahara, N. Tsubouchi, Y. Horino, and J.S. Williams, Nucl. Instrum. Meth. Phys. Res. B **148**, 370 (1999).
4. A. Kinomura, J.S. Williams, and K. Fujii, Phys. Rev. B **59**, 15214 (1999).
5. P. Pichler, *Intrinsic Point Defects, Impurities, and their Diffusion in Silicon* (Springer, Wien, 2004).
6. P. Spiewak and K.J. Kurzydlovski, Phys. Rev. B **88**, 195204 (2013)
7. M.J. Beck, L.Tsetseris, and S.T. Pantelides, Phys. Rev. Lett. **99**, 215503 (2007).
8. B. J. Pawlak, N. E. B. Cowern, C. Ahn, W. Vandervorst, R. Gwilliam and J. G. M. van Berkum, Appl. Phys. Lett. **105**, 221603 (2014).
9. G.D. Watkins, in *Radiation Damage in Semiconductors* (Dunod, Paris, 1964), p. 97.
10. E.J.H. Collart *et al.*, Nucl. Instrum. Meth. Phys. Res. B **139**, 98 (1998).
11. E. Napolitani *et al.*, Phys. Rev. Lett. **93**, 055901 (2004).
12. N.E.B. Cowern, G. Mannino, P.A. Stolk, F. Roozeboom, H.G.A. Huizing, J.G.M. van Berkum, F. Cristiano, A. Claverie, and M. Jaraíz, Mat. Sci. Semicond. Proc. **2**, 369 (1999).
13. C.F. Chien and W.W. Soroka, J. Sound Vibr. **43**, 9 (1975).
14. A. Yurtsever and A. Zewail, Proc. Nat. Acad. Sci. **108**, 3152 (2011) (Supporting Information)
15. B.L. Zink, R. Pietri, and F. Hellman, Phys. Rev. Lett. **96**, 055902 (2006).
16. P. Hess and A.M. Lomonosov, Ultrasonics **50**, 167 (2010)
17. F. Kail et al., Physica Status Solidi (*RRL*) **5**, 361 (2011).
18. A. Polman, S. Roorda, P.A. Stolk, and W.C. Sinke, J. Crystal Growth **108**, 114 (1991).
19. S. Hayashi, Y. Fujita, T. Kamikura, K. Sakaike, M. Akazawa, M. Ikeda, and S. Higashi, Jap. J. Appl. Phys. **52**, 05EE02 (2013).

**Figure captions**

Fig. 1.
Atomic profiles of 20 keV, $3\times10^{15}/cm^2$ B implanted into a/c-Si bilayers at various temperatures. The shift of the B pile-up position with temperature is caused by ion beam induced regrowth. The $T$ = 77 K SIMS profile has been offset to compensate for the presence of an ice layer on the a-Si surface after cooling the sample for implantation. The ice layer thickness is equivalent in terms of stopping power to 15 nm of Si. This has shifted the a/c interface position 15 nm to the right. It appears as a weak discontinuity in the slope of the 77 K curve relative to the 293 K curve, starting at a depth of 110 nm. We estimate that this corresponds to an a/c interface depth of (95 ± 3) nm in the original sample implanted at 77 K. The error is given relative to the common depth calibration for all the profiles.

Fig. 2.
A/c interface depths corresponding to the data shown in Fig. 1, plotted as a function of inverse temperature. Symbols are experimental data and the curve is an Arrhenius fit to the data at room temperature and above, using a single activation energy of 0.07 eV. The curve extrapolates perfectly to fit the data at 77 K (symbol at far right).

Fig. 3.
The shift in a/c interface position arising from beam-induced SPER, presented as an Arrhenius plot. The data are best fitted by a straight line with activation energy 0.07 eV and a prefactor of 71 nm. In the high-temperature limit this corresponds to incorporation of 118 'amorphous' Si atoms into the c-Si region per single implanted B ion, or several hundred times the number of B ions stopped in the c-Si region.

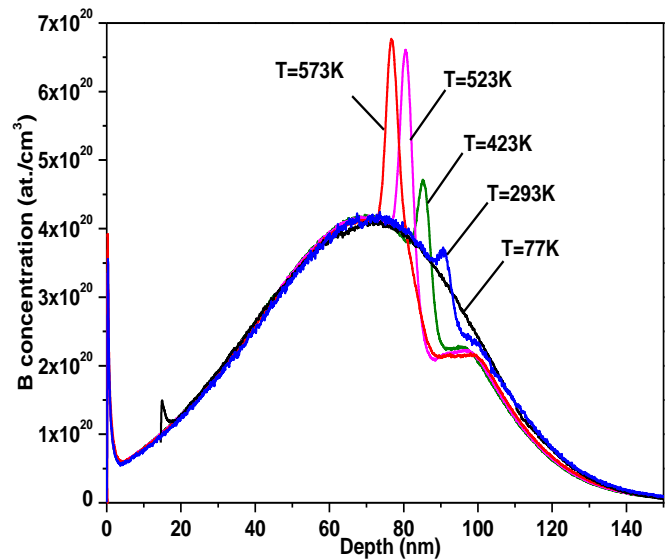

Fig. 1.
Atomic profiles of 20 keV, $3\times10^{15}/cm^2$ B implanted into a/c-Si bilayers at various temperatures. The shift of the B pile-up position with temperature is caused by ion beam induced regrowth. The $T = 77$ K SIMS profile has been offset to compensate for the presence of an ice layer on the a-Si surface after cooling the sample for implantation. The ice layer thickness is equivalent in terms of stopping power to 15 nm of Si. This has shifted the a/c interface position 15 nm to the right. It appears as a weak discontinuity in the slope of the 77 K curve relative to the 293 K curve, starting at a depth of 110 nm. We estimate that this corresponds to an a/c interface depth of $(95 \pm 3)$ nm in the original sample implanted at 77 K. The error is given relative to the common depth calibration for all the profiles.

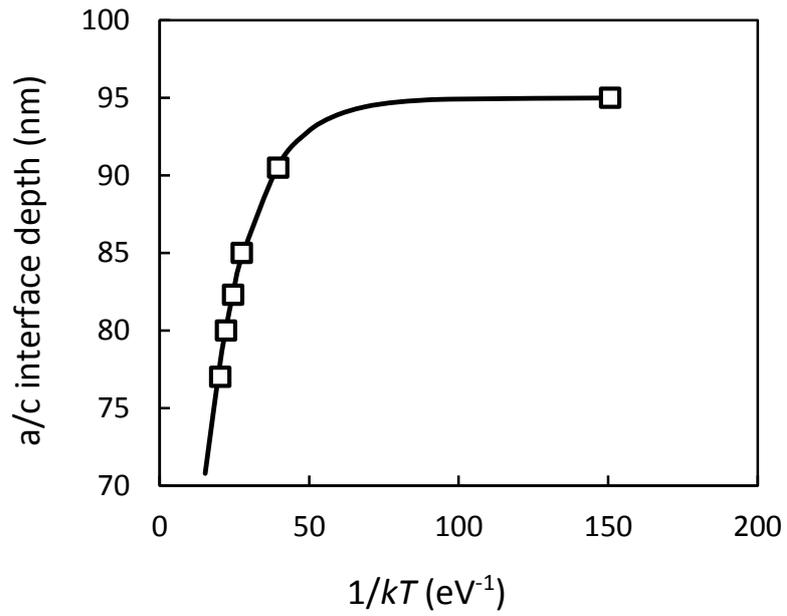

Fig. 2.
A/c interface depths corresponding to the data shown in Fig. 1, plotted as a function of inverse temperature. Symbols are experimental data and the curve is an Arrhenius fit to the data at room temperature and above, using a single activation energy of 0.07 eV. The curve extrapolates perfectly to fit the data at 77 K (symbol at far right), confirming there is no significant regrowth at 77K.

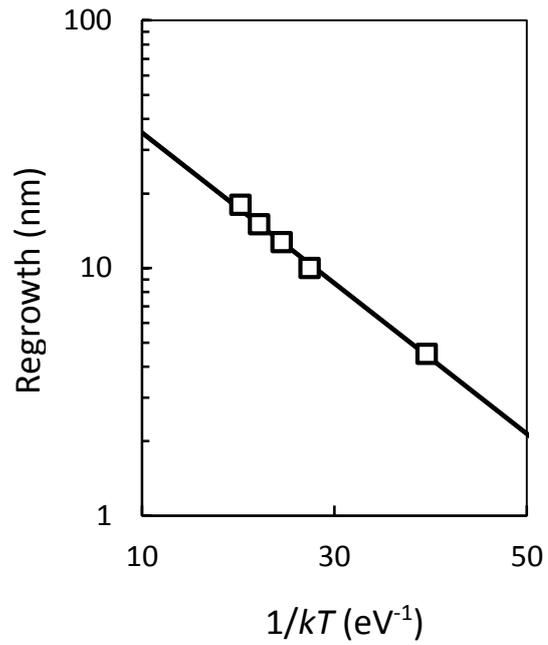

Fig. 3.
The shift in a/c interface position arising from beam-induced SPER, presented as an Arrhenius plot. The data are best fitted by a straight line with activation energy 0.07 eV and a prefactor of 71 nm, corresponding to a prefactor / dose ratio of 2.36 nm / $10^{14}$ cm$^{-2}$. In the high-temperature limit this corresponds to incorporation of 118 'amorphous' Si atoms into the c-Si region per single implanted B ion, or ~ $10^3$ times the number of B ions stopped in the c-Si region.